\documentclass[aps,prb,twocolumn,superscriptaddress,citeautoscript]{revtex4-2}

\setcitestyle{super}

\usepackage{epsfig}
\usepackage{amsfonts}
\usepackage{graphics}
\usepackage{longtable}
\usepackage{epsfig}
\usepackage{amsmath}
\usepackage{float}
\usepackage{placeins}
\usepackage{color,soul}
\usepackage{multirow,tabularx}
\usepackage{titlesec}
\usepackage{lineno}
 
\titleformat{\section}
{\normalfont\Large\bfseries}{\thesection}{1em}{}

\newcommand*{\citen}[1]{%
	\begingroup
	\romannumeral-`\x 
	\setcitestyle{numbers}%
	\cite{#1}%
	\endgroup   
}

\begin{document}
\bibliographystyle{apsrev}
\title{Embedding material graphs using the electron-ion potential: application to material fracture}
\author{Sherif Abdulkader Tawfik$^{\dagger}$}
\email{s.abbas@deakin.edu.au}
\affiliation{Applied Artificial Intelligence Institute, Deakin University, Geelong, Victoria 3216, Australia.}
\affiliation{ARC Centre of Excellence in Exciton Science, Deakin University, Geelong, Victoria 3216, Australia.}

\author{Tri Minh Nguyen$^{\dagger}$}
\email{tri.nguyen1@deakin.edu.au}
\affiliation{Applied Artificial Intelligence Institute, Deakin University, Geelong, Victoria 3216, Australia.}

\author{Salvy P. Russo}
\affiliation{ARC Centre of Excellence in Exciton Science, School of Science, RMIT University, Melbourne 3000, Australia.}

\author{Truyen Tran}
\affiliation{Applied Artificial Intelligence Institute, Deakin University, Geelong, Victoria 3216, Australia.}

\author{Sunil Gupta}
\affiliation{Applied Artificial Intelligence Institute, Deakin University, Geelong, Victoria 3216, Australia.}

\author{Svetha Venkatesh}
\email{svetha.venkatesh@deakin.edu.au}
\affiliation{Applied Artificial Intelligence Institute, Deakin University, Geelong, Victoria 3216, Australia.}\

\begin{abstract}
	

	
	At the heart of the flourishing field of machine learning potentials are graph neural networks, where deep learning is interwoven with physics-informed machine learning (PIML) architectures. Various PIML models, upon training with density functional theory (DFT) material structure-property datasets, have achieved unprecedented prediction accuracy for a range of molecular and material properties. A critical component in the learned graph representation of crystal structures in PIMLs is how the various fragments of the structure's graph are embedded in a neural network. Several of the state-of-art PIML models apply spherical harmonic functions. Such functions are based on the assumption that DFT computes the Coulomb potential of atom-atom interactions. However, DFT does not directly compute such potentials, but integrates the electron-atom potentials. We introduce the direct integration of the external potential (DIEP) methods which more faithfully reflects that actual computational workflow in DFT. DIEP integrates the external (electron-atom) potential and uses these quantities to embed the structure graph into a deep learning model. We demonstrate the enhanced accuracy of the DIEP model in predicting the energies of pristine and defective materials. By training DIEP to predict the potential energy surface, we show the ability of the model in predicting the onset of fracture of pristine and defective carbon nanotubes.
	
	$^{\dagger}$ These authors contributed equally to this work.
\end{abstract}

\maketitle

\section{Introduction}



In the last few years, machine learning (ML) potential models have been amassing an unprecedented number of contributions from interdisciplinary research teams worldwide. The capabilities of these models rapidly expanded into various material science applications, promising a future where highly accurate quantum material science computations can be performed at the cost of classical molecular dynamics, if not at a lower cost. The accuracy and generalisability of the models have been empowered by two key factors: the emergence of graph neutral network (GNN)\cite{baskin_neural_1997} models that superseded standard ML models in accuracy and complexity,\cite{zhang_artificial_nodate} and the abundance of a massive amount of quantum mechanically-computed material data in online databases. Over the last few years, the online databases, such as Materials Project (MP),\cite{jain_commentary_2013} JARVIS,\cite{choudhary_joint_2020} AFLOW,\cite{curtarolo_aflow_2012} Open Quantum Mechanical Database (OQMD),\cite{kirklin_open_2015} and others, have availed more than 3 million DFT-computed structures, and have been part of a standard benchmarking workflow for new ML models. GNNs have further been improved by the introduction of physical laws within the fabric of the neural network, establishing what is commonly known as physics-informed machine learning (PIML) models, as was demonstrated in DimeNet,\cite{gasteiger_directional_2022} its derivative M3GNET\cite{chen_universal_2022} and MACE\cite{batatia_mace_2023} for property prediction and structure discovery. These models have explicitly incorporated physics-based representations for the atomic structure by transforming doublets (atom-atom pairs) using a radial basis function (RBF), and triplets (groups of three atoms within a sphere of a given radius) using spherical harmonics. The utilization of spherical harmonics stemmed from the analytical structure of the wave function, which is the solution of the Schr\"{o}dinger equation that DFT aims to approximate. Trained on over 133k samples, the M3GNET achieved a mean absolute error (MAE) of 20 meV/atom for predicting the formation energy ($E_f$) of the test set samples. However, as will be shown in this work, the model struggles with the prediction of the properties of defective crystals, and in predicting the onset of carbon nanotube fracture.

\begin{figure*}[t]
	\begin{center}	
		\includegraphics[width=\textwidth]{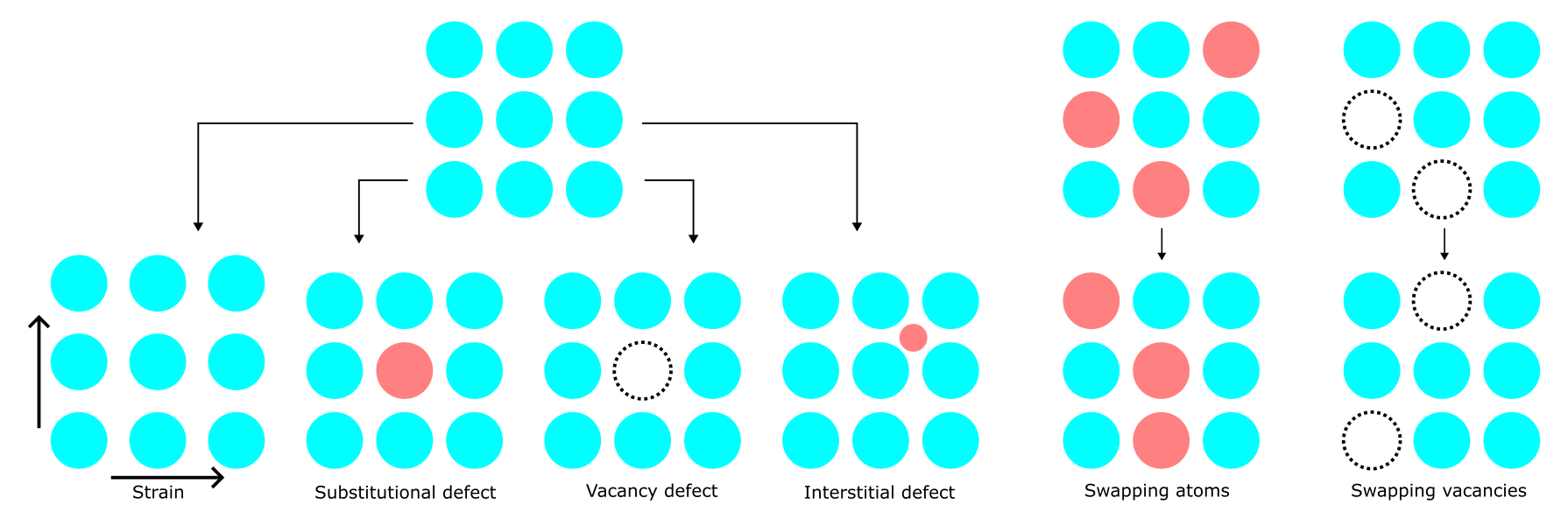}
	\end{center}
	\caption{\label{fig_defects}The types of crystal deformation that are examined in this work: strain deformation, where the crystal lattice size if modified to reflect the application of either a stretching or compressive force on the structure; substitutional defect, where one of the atoms is substituted with another; vacancy defect, where an atom is removed from the crystal leaving a vacancy; interstitial defect, where an atom is introduced into a void in the crystal; swapping of atoms, where atoms within the crystal exchange positions; swapping vacancies, where vacancies, rather than atoms, are exchanged.}
\end{figure*}

We propose a novel physics-informed ML framework for accurate prediction of both pristine and imperfect crystal materials. Called DIEP, it directly integrates the external potential of a structure, and it was implemented on the codebase of M3GNET. This direct integration is a critical correction to the physics in M3GNET, and we show its value in discerning the impact of material defects. In particular, DIEP is able to more accurately predict the total energy per atom of a defective system, as well as the structural changes that result from the presence of a defect in a material. We demonstrate these two merits of the DIEP model by conducting two learning tasks using datasets of crystalline systems: in task 1, we train the models to predict the total energy per atom and demonstrate the accuracy of DIEP in predicting the total energy per atom for 6 classes of material imperfections, which are displayed in Figure \ref{fig_defects}. We further interrogate the models on common defects of diamond, and show that the accuracy of DIEP exceeds that of the trained M3GNET model in most of the test cases. In task 2, we train a potential model that predicts both the energy per atom and the atomic forces, which amount to the prediction of the potential energy surface (PES). We show the ability of DIEP in (1) reproducing the ground state crystal structure of a number of binary materials, and (2) in computing the fracture strain of a large carbon nanotube structure due to the presence of carbon vacancy defects, while the M3GNET model does not predict rupture of the CNT at excessive elongation strains.



\section{Improving the physics}

The PIML in M3GNET is based on embedding the graph nodes (including the atom-atom pairs, or bonds, and the three-atom structures, or triplets) into the neural network through a layer that calculates DFT-related properties, as was initially proposed in the DimeNET model. The DFT-related properties assume that the atoms in the structure are directly coupled by a Coulomb potential in a simplified Schr\"{o}dinger equation, and embed the nodes by using the solutions of that equation as functions of the bonds and triplets in the structure. However, the direct atom-atom interaction emulated by the M3GNET is in fact neglected in typical DFT calculations. This is because, based on the Born-Oppenheimer approximation, the positions of atoms are treated as fixed. In this work, we introduce an alternative PIML approach that is based on the numerical integration of the external potential which is directly relevant to DFT. We embed the graph nodes using an integration layer, in which the atom-electron potential, rather than the atom-atom potential, is computed. Computing an integration for atom-electron interaction is advantageous because the atom-electron interaction uniquely determines the ground state DFT electron density $\rho(\textbf{r})$ (a direct implication of the Honenberg-Kohn theorem (\cite{PhysRev.136.B864})), where $\rho(\textbf{r})$ is the function that enables the determination of the ground state total energy. Our method computes a simplified form for this term for the graph's bonds and triplets on a two-dimensional mesh, and passes these atom-electron interaction messages into the GNN. 

The total DFT energy is calculated by summing the kinetic energy $T[\rho(\textbf{r})]$, atom-electron energy (or external energy) $E_{ext}[\rho(\textbf{r})]$, the Hartree energy $E_{H}[\rho(\textbf{r})]$, and the exchange correlation energy $E_{XC}[\rho(\textbf{r})]$, which are all functionals of the electron density $\rho(\textbf{r})$:

\begin{equation}
	E_{T}[\rho(\textbf{r})] = T[\rho(\textbf{r})] + E_{ext}[\rho(\textbf{r})] + E_{H}[\rho(\textbf{r})] + E_{XC}[\rho(\textbf{r})]
	\label{eq1}
\end{equation}

\noindent where the terms are given by

\begin{eqnarray}
	\label{eq_hartree}
	T[\rho(\textbf{r})] &=& -\frac{1}{2}\sum_{n\textbf{k}}^{N} \int d\textbf{r} \phi^{*}(\textbf{r}) \nabla^2 \phi(\textbf{r}) \label{eq_kineticenergy} \\ \nonumber
	E_{ext}[\rho(\textbf{r})] &= &\int d\textbf{r} \sum_{N}^{n}\frac{\rho(\textbf{r})Z_n}{|\textbf{r} - \textbf{R}_n|} \label{eq_ext} \\ \nonumber
	E_{H}[\rho(\textbf{r})] &=&\frac{1}{2} \int \int  d\textbf{r} d\textbf{r}^{'}\frac{\rho(\textbf{r})\rho(\textbf{r}^{'})}{|\textbf{r} - \textbf{r}^{'}|} \\ \nonumber
\end{eqnarray}

\noindent The value of $E_{T}$ must be minimized with respect to the given atomic structure, and therefore one must find a density function $\rho(\textbf{r})$ that will yield such minimum energy. Given the definition of the electron density in terms of the fictitious electronic orbital functions $\phi_i(\textbf{r})$

\[
\rho(\textbf{r}) = \sum_{i}^{N_e} \phi^{*}_i(\textbf{r})\phi_i(\textbf{r})
\]

\noindent where $\phi^{*}_i(\textbf{r})$ is the complex conjugate of $\phi_i(\textbf{r})$, the minimization problem is solved by applying the variational principle to $E_{T}$ with respect to the orbital functions $\phi_i(\textbf{r})$. This yields the Kohn-Sham equations,

\begin{equation}
	H_{KS} \phi_i(\textbf{r}) = E_i\phi_i(\textbf{r})
\end{equation}

\noindent where the Kohn-Sham Hamiltonian $H_{KS}$ is given by

\begin{eqnarray}
	\label{eq_KS}
	H_{KS} &=& -\frac{1}{2}\nabla^2 + V_{ext}(\textbf{r}) + V_H(\textbf{r}) \\  \nonumber
	V_{ext}(\textbf{r}) &=&  \sum_{N}^{n}\frac{Z_n}{|\textbf{r} - \textbf{R}_n|} \\ \nonumber
	V_H(\textbf{r}) &=& \frac{1}{2} \int  d\textbf{r}^{'}\frac{\rho(\textbf{r}^{'})}{|\textbf{r} - \textbf{r}^{'}|}\\ \nonumber
\end{eqnarray}

\noindent The only term in $H_{KS}$ that is directly dependent on the positions of the atoms is $V_{ext}(\textbf{r})$. Likewise, the only term in $E_{T}[\rho(\textbf{r})]$ that is directly dependent on the positions of the atoms is $E_{ext}[\rho(\textbf{r})]$. \textit{These terms, however, do not use direct atom-atom distances, contrary to the assumption made in the M3GNET and DimeNet models}. 

In a GNN, structural data are generated for the input structure into a graph object with nodes and edges. For a given structure, nodes can represent information about atoms at the nodes' spacial positions ${\textbf{R}_i}$, and edges can represent the bonds between the atoms. The structure of the graph, as well as the count of nodes and edges, is different for each structure. In a GNN, the NN learns the graph connection between each node and the neighboring nodes \textit{iteratively}. Following the extraction of bonds and triplets from the graphical representation of the input crystal structure, which is routinely performed by common GNN-based packages, DIEP performs the following standardization transformation on bonds and triplets, which is depicted in Figure \ref{fig_method}:

\begin{figure*}[t]
	\begin{center}	
		\includegraphics[width=\textwidth]{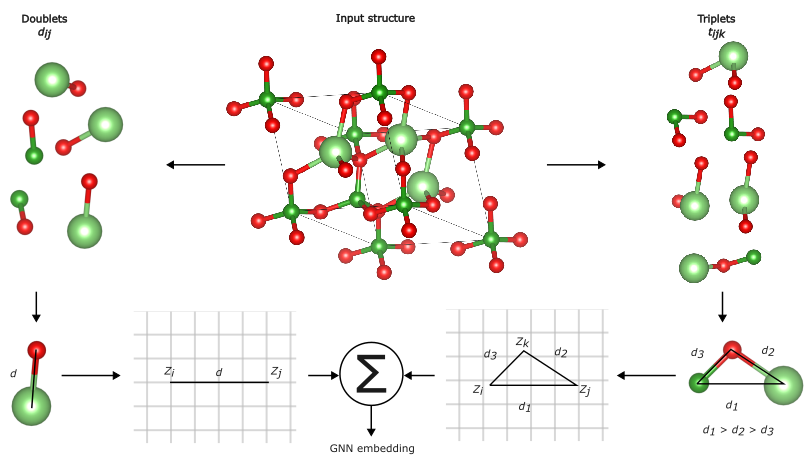}
	\end{center}
	\caption{\label{fig_method} \textbf{Outline of the DIEP method and characteristics of the TinyUnitCells dataset.} The input crystal structure is decomposed into its constituent bonds $d_{ij}$ and triplets $t_{ijk}$. Then, each bond is transformed into a line that is centred in a sparse 2D mesh, and each triplet is transformed into a triangle with ordered side lengths, and is also centred in a sparse 2D mesh. The DIEP integration is applied to each $d_{ij}$ and $t_{ijk}$ in their respective 2D meshes, and the value of each integration is used to embed the $d_{ij}$ and $t_{ijk}$ of the input structure. (b) Comparison between the number of triplets extracted for materials in the TinyUnitCells dataset ($\leq 10$ atoms), and defective materials with $> 10$ atoms, showing the large difference in the distribution of geometric information between the two sets of materials. (c) Periodic table displaying a heatmap of the frequency of occurrence of each element in the periodic table in the TinyUnitCells dataset.}
\end{figure*}

\begin{itemize}
	\item for a bond between atoms with atomic numbers $Z_i$ and $Z_j$, positions $\textbf{R}_i$ and $\textbf{R}_j$, the line joining the points $\textbf{R}_i$ and $\textbf{R}_j$ is transformed into the line joining the two points $-\frac{d_{ij}}{2}$ and $\frac{d_{ij}}{2}$, where $d_{ij}$ is the distance between  $\textbf{R}_i$ and $\textbf{R}_j$;
	\item for a triplet involving three atoms with atomic numbers $Z_i$, $Z_j$ and $Z_k$, positions $\textbf{R}_i$, $\textbf{R}_j$ and $\textbf{R}_k$, the triangle $\textbf{R}_i$ --- $\textbf{R}_j$ --- $\textbf{R}_k$ is transformed to $\textbf{R}_a$ --- $\textbf{R}_b$ --- $\textbf{R}_c$ such that $\textbf{R}_a$ --- $\textbf{R}_b$ is the longest side, followed by  $\textbf{R}_b$ --- $\textbf{R}_c$ then  $\textbf{R}_a$ --- $\textbf{R}_c$.
\end{itemize}

\noindent For a transformed bond, identifying either atoms as $Z_i$ or $Z_j$ is the same, whereas for a triplet, we set the identities of the atoms in the transformed triangle based on the intersection between the identities of the atom pairs in the bonds of the triangle. This transformation ensures that the transformed structure is invariant to permutation of the identities of atoms.

Both the transformed bond and triangle are then centred within a two-dimensional grid where the $x$- and $y$ axes range from $-L$ to $+L$ (in units of~\AA), with step size $\Delta l$. The transformation of the triangle $\textbf{R}_i$ --- $\textbf{R}_j$ --- $\textbf{R}_k$ ensures that any three atoms at positions $\textbf{R}_i$, $\textbf{R}_j$ and $\textbf{R}_k$ will have a unique value for $E_{ijk}$. 

The 2D grid in DIEP is a simplification for the actual 3D grid in which the external potential is integrated; it is only a 2D ``slice'' of the 3D box. The points on the mesh represent the coordinates $\textbf{r}$ of the electronic density function $\rho(\textbf{r})$. The representation of a triplet as a triangle on a mesh is an explicit incorporation of three-body interactions that combines the properties of the atoms (atomic numbers \textbf{$Z_i$}) with their positions in a natural way, rather than embedding the triplet using bond angles that are separated from the atomic numbers as in M3GNET.

To embed a transformed bond, we compute the following terms, which are based on the external energy term $E_{ext}[\rho(\textbf{r})]$ in Eq. \ref{eq_ext}:

\begin{equation}
	\label{eq_e_ij}
	E_{ij,mn}  = \Delta l^2\rho(\textbf{r}_{mn})\left( \frac{Z_i}{|\textbf{r}_{mn}-\frac{d_{ij}}{2}|} + \frac{Z_j}{|\textbf{r}_{mn}+\frac{d_{ij}}{2}|}\right) \quad,
\end{equation}

\noindent whereas for triplets, the following quantities are computed:

\begin{eqnarray}
	\label{eq_e_ijk}
	E_{ijk,mn}  &=&\Delta l^2\rho(\textbf{r}_{mn}) \\ \nonumber
	&\times&\left(\frac{Z_i}{|\textbf{r}_{mn}-\textbf{R}_{a}|} + \frac{Z_j}{|\textbf{r}_{mn}-\textbf{R}_{b}|} + \frac{Z_k}{|\textbf{r}_{mn}-\textbf{R}_{c}|}\right)\quad.\\ \nonumber
\end{eqnarray}

\noindent where $\rho(r_{mn})$ represents a simple analytical form for the electron density, which is given by

\begin{equation}
	\rho(\textbf{r}_{mn}) = \sum_{i}^{N}e^{-\frac{|\textbf{r}_{mn}-\textbf{R}_{i}|^2}{\sigma}}
\end{equation}

\noindent and $m$ and $n$ are indices from the position of a point in the 2D grid, and the grid is of size $L\times L$. The function $\rho(\textbf{r})$ is the optimization target in DFT, and therefore its relationship with the atomic structure is highly complicated. One can use a simplified or trained\cite{cuevas-zuviria_analytical_2020} analytical form for $\rho(\textbf{r})$. \textcolor{black}{However, we choose a simplified representation: a summation of gaussian functions centred at the atomic positions. This representation is based on the trainable expression (Eqs. 1-3) in Ref. \cite{cuevas-zuviria_analytical_2020}, in which the electron density is expanded as sum of gaussian functions with trainable coefficients.}

\textcolor{black}{While this charge density representation is not representative for a large number of bonds, particularly in charge transfer situations where one of the atom loses a significant portion of its charge density (such as in the case of transition metals), it is still a reasonable assumption for the broad variety of bonds, as well as being computationally efficient. An improvement to the charge density representation would either use parametrised expressions such the one in Ref. \cite{cuevas-zuviria_analytical_2020}, or to express the charge density as a sum of gaussians with learnable parameters. Those learnable parameters would then be tuned as part of the entire training cycles of the DIEP model. Even though these trained 2D fragment charge densities will not be as complete as the densities that are obtained from models trained on actual densities, such as Refs. \cite{sunshine_chemical_2023,gong_predicting_2019,jorgensen_deepdft_2020}, they would be expected to model the densities of their respective fragments. Exploring alternative strategies to represent $\rho(\textbf{r})$ is currently in progress.}

Following the computation of the quantities $E_{ij,mn}$ and $E_{ijk,mn}$, these quantities are then used to embed the bonds $ij$ and triplets $ijk$ into the neural network. \textcolor{black}{This by performing the following numerical integration across the grid:} 

\begin{eqnarray}
	\label{eq_e_ij_sum}
	E_{ij}  &=& \sum^{L}_{m=1}\sum^{L}_{n=1}E_{ij,mn} \quad,\\  \nonumber
	E_{ijk}  &=& \sum^{L}_{m=1}\sum^{L}_{n=1}E_{ijk,mn}
\end{eqnarray}

\textcolor{black}{The quantities $E_{ijk}$ and $E_{ij}$ are then fed into the GNN.} These quantities replace the 2D representation quantities $a^{(kj,ji)}_{SBF}$ and the distance representation quantities $e^{(ji)}_{RBF}$ in the DimeNet embedding layer (Figure 4 of Ref. \citen{gasteiger_directional_2022}). The total number of trainable parameters in the DIEP model is 279,837, whereas it is 288,157 in the M3GNET model. \textcolor{black}{Hence, an improvement in the DIEP model is not due to introducing additional learnable parameters.} The DIEP hyperparameters include: the resolution of the grid (the values of $L$ and $\Delta l$), the choice of using the external potential or external energy terms, whether to use the entire grid points or to use the summation of terms, and the value of $\sigma$ in the exponential formula for $\rho(\textbf{r})$.



Next, To examine the predictive power of DIEP compared with M3GNET, we perform two training tasks. In the first task, we train DIEP and M3GNET models on a subset of stable materials from MP, and then challenge the generalisability the two models on deformed materials which are described in Figure \ref{fig_results_defects}. In the second task, we train the M3GNET and DIEP models on a dataset of structural optimisation trajectories (the MPF.2021.2.8 dataset used in Ref.~\citen{chen_universal_2022}) and compare the accuracy of predicting the total energy per atom and the atomic forces. 

\section{Results and discussion}
\subsection{Prediction of the energy of defective materials}

\noindent\textbf{Defective materials} We design material deformations with structural features that are significantly different from the materials in TinyUnitCells. We consider deformations that are often encountered in routine material science investigations, as well as those generated in the course of material discovery processes. The classes of deformations we examine are:

\begin{itemize}
	\item \textcolor{black}{\textit{Random strains}: For a sample of unit cells with a maximum of 4 atoms, a uniaxial strain is applied, where we randomly pic a strain value from the set of percentage strain values $\pm 1$\%, $\pm 2$\%, $\pm 3$\%, $\pm 4$\% and $\pm 5$\% along the $a$ lattice direction. For another sample of unit cells with a maximum of 4 atoms, we applied random triaxial strains, where we randomly pic a strain value from the set of percentage strain values $\pm 1$\%, $\pm 2$\%, $\pm 3$\%, $\pm 4$\% and $\pm 5$\% along each of the axial directions.}
	\item \textit{Single-site defects}: For a sample of unit cells with a maximum of 4 atoms, a point defect is created in $2\times 2 \times 1$ supercells by either removing an atom, or substituting an atom with another, such that the two atoms are in the same group of the periodic table. This ensures the same number of valence electrons in the unit cell and avoid the complexity of dealing with potentially charged defects.
	\item \textit{Swapping of atomic positions}: This disorder naturally occurs in high entropy alloys, where the distribution of the atoms within the lattice structure is rather stochastic.\cite{batchelor_high-entropy_2019}
	\item \textit{Multi-site defects}: A permutation of multiple vacancies is created in a supercell. In some cases, the permutations do not result in significant changes in the total energy. In this case, this class of deformations examines the \textit{false positives}: whether the ML model will overestimate the energy cost of a small geometric change.

	\item \textit{Interstitial defects}: Hydrogen atoms are introduced into interstitial sites of metals and perovskites.

\end{itemize}

\begin{figure}[t]
	\begin{center}	
		\includegraphics[width=0.5\textwidth]{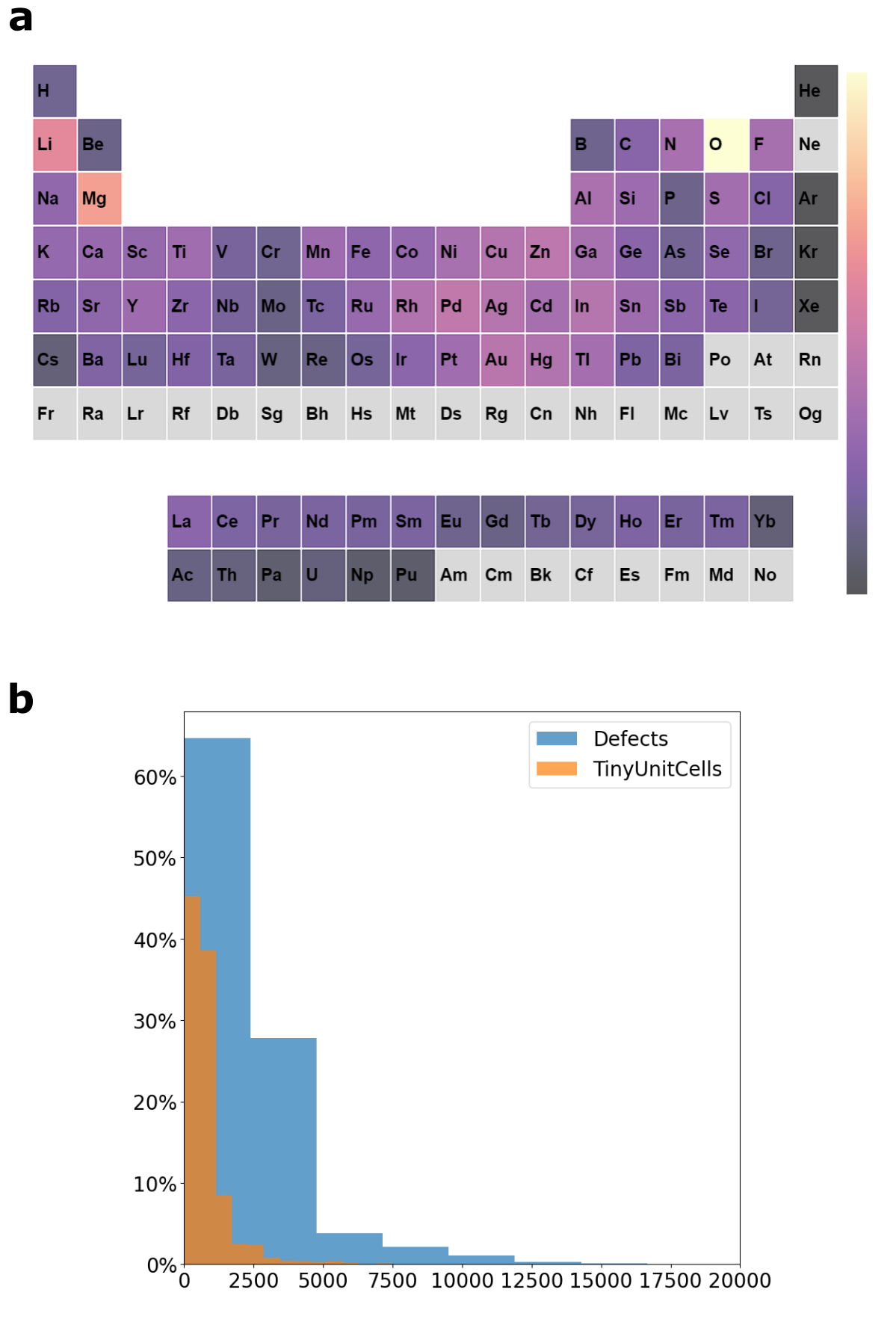}
	\end{center}
	\caption{\label{fig_dataset} \textbf{Dataset characteristics.} (a) Periodic table displaying a heatmap of the frequency of occurrence of each element in the periodic table in the TinyUnitCells dataset. (b) Comparison between the number of triplets extracted for materials in the TinyUnitCells dataset ($\leq 10$ atoms), and defective materials with $> 10$ atoms, showing the large difference in the distribution of geometric information between the two sets of materials. }
\end{figure}

To examine the ability of DIEP and M3GNET in predicting the energies of the above defects, we first train both models on a sample of pristine materials from MP with up to 10 atoms in the unit cell, TinyUnitCells (version 1). The composition of this dataset is: 10k materials with up to 4 atoms/unit cell, 1k with 5 or 6 atoms/unit cell, and 1k materials with $7-10$ atoms/unit cell. The representation of each of the elements in TinyUnitCells is displayed in Figure \ref{fig_dataset}a. There are 88 elements represented in the dataset (out of 89 elements that are represented in MP), the most frequently occurring elements are O, Li and Mg. The large representation of O in TinyUnitCells is similar to the case of the larger dataset MPF.2021.2.8.\cite{chen_universal_2022}

\begin{figure}[H]
	\begin{center}	
		\includegraphics[width=0.5\textwidth]{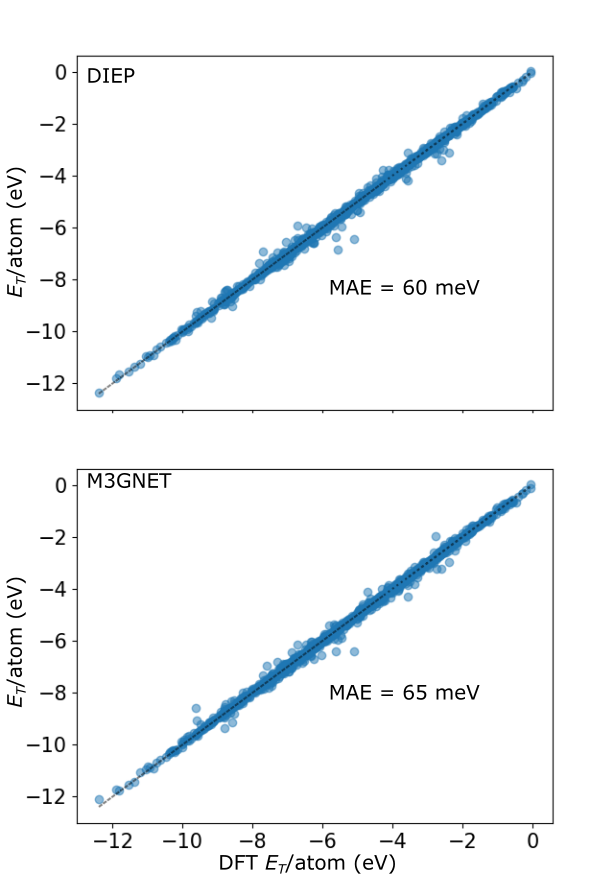}
	\end{center}
	
	\caption{\label{fig_results_correlations} Correlation plots for the ML prediction of the total energy/atom for the structures in TinyUnitCells using (a) DIEP and (b) M3GNET.}
\end{figure}

\noindent\textbf{TinyUnitCells results} We have re-computed the DFT structural optimisation for TinyUnitCells using standardised DFT input parameters (details in the Methods section). The size of unit cells in TinyUnitCells are chosen in order to bias the GNN on learning features of small unit cells, and hence to test the ability of the method to generalise to larger structures, particularly those supercells with a small concentration of defects. The histogram in Figure \ref{fig_dataset}b illustrates the scale of such difference in terms of the number triplets extracted from each structure in the dataset. Small unit cells (such as those in TinyUnitCells) possess much lower triplets than larger structures, such as the defect structures examined here. The 80\%-10\%-10\% split is also applied for the TinyUnitCells dataset. DIEP exceeds the accuracy of M3GNET in predicting the total energy/atom of the test set, as displayed in Figure \ref{fig_results_correlations}b. Note that the MAE values are higher than those obtained for the JARVIS dataset because the size of TinyUnitCells is nearly one-fifth that of JARVIS. After establishing the accuracy of training the DIEP models for the pristine materials in TinyUnitCells, we demonstrate its accuracy for predicting the defective crystals.

\noindent\textbf{Prediction of the total energy pert atom} The results of running the two models on the 6 classes of deformed materials is displayed in Table \ref{tab_results_deformations}. DIEP performs better than M3GNET in reproducing the total energy/atom for most of the defect classes considered.

\begin{table*}[t]
	\caption{The mean absolute error (MAE) in meV/atom for predicting the total energy/atom of the deformed materials. The atomic structures of all the deformation classes have been relaxed except for the ``unoptimised uniaxial'' strains.}
	\label{tab_results_deformations}
	\begin{center}
			\begin{tabular}{|ll|lll|}
				\hline
				\textbf{Deformation} & & \textbf{Dataset size}&	\textbf{DIEP MAE (meV/atom)} &	\textbf{M3GNET MAE (meV/atom)}\\
				\hline
				
				\multirow{3}{*}{Strains} &Triaxial &404 & \textbf{90}&	94 \\

				 &Optimised uniaxial & 929& \textbf{59}&	65 \\
 				 &Unoptimised uniaxial & 943& \textbf{61}&	66 \\
				\hline
				
				\multirow{2}{*}{Single-site defects} & Vacancies & 213 & \textbf{78}& 79	 \\
				 & Substitutions & 196 & \textbf{108}& 120	 \\
				\hline
				
				\multirow{2}{*}{Atomic swaps}  & PtPdIrRh & 257 & \textbf{41}	& 55 \\
			 &CoCrFeNi& 263& 90&\textbf{45} \\
				&FeNi& 123&	328&\textbf{287} \\
				&PtIr&268 &	\textbf{41}&43 \\
				\hline
				
				\multirow{2}{*}{Multi-vacancy swaps} & LiCoO$_2$&53 &\textbf{53} &112 \\
			 &LiAlO$_2$ &53 &96&\textbf{11}  \\
			 &LiFePO$_4$ &70 & \textbf{195}&231	 \\
				\hline
				
				\multirow{2}{*}{Single interstitial H defect} & Metals & 246& 97	&\textbf{90} \\
			 & Perovskites&141 &\textbf{91}	&101 \\
				\hline
				
				\multirow{4}{*}{Multiple interstitial H defects} & Zn & 18& \textbf{48}	&103 \\
				 & Ni&14 &\textbf{24}	&35 \\
				 & Fe&10 &\textbf{136}	&180 \\
				  & Pt&10 &\textbf{56}	&144 \\
			\hline
			\end{tabular}
	\end{center}
\end{table*}

DIEP outperforms M3GNET in predicting the energies of strained unit cells\textcolor{black}{, whether the strain is triaxial, uniaxial optimised or uniaxial unoptimised, as shown in Table \ref{tab_results_deformations}. We further examine the optimised uniaxial strained structures by observing the errors obtained for the strains $\pm 1$\%, $\pm 2$\%, $\pm 3$\%, $\pm 4$\%, and $\pm 5$\%. The MAE for the DIEP model for each group of strains gradually increase with strain, as expected: 37 meV/atom, 45 meV/atom, 56 meV/atom, 68 meV/atom and 94 meV/atom. Except for the $\pm 4$\% strain, these errors are consistently lower than those obtained using the trained M3GNET model: 44 meV/atom, 53 meV/atom, 62 meV/atom, 67 meV/atom and 101 meV/atom.}

For single-site defects, DIEP is only 7 meV/atom better than M3GNET. Both models can accurately predict substitutional defects, where an atom is replaced by another atom that is one row above or below it in the periodic table (such as F replaced with Cl). They both struggle to predict the energy of substitutional defects where a small atom is substituted with a larger one that is 2 or more rows lower in the periodic table. For example, the DFT-calculated total energy/atom for BN (\texttt{mp-1580}) with a N atom substituted with Bi is $-7.90$ eV/atom, but DIEP predicted a value of $-9.87$ eV/atom and M3GNET predicted $-9.74$ eV/atom. Vacancy defects are generally accurately predicted.

For alloy structures with swapped atomic positions, atomic swapping leads to mild changes in the total energy/atom in the PtPdIrRh, CoCrFeNi and PtIr alloy systems, and significant changes in the FeNi alloy system. The range of energies (difference between the highest and lowest values) in these alloy systems is 44 meV/atom, 73 meV/atom, 21 meV/atom and 354 meV/atom, respectively. Such spread in the data affected the predictive accuracy of both models. We also examine the swapping of multiple vacancy sites in the cathode materials LiCoO$_2$ and LiFePO$_4$, and LiAlO$_2$, a cathode coating material. Lithium diffusion is key to the function of these materials, and it involves the presence of various defect configurations at any instance of time. We swap 3 defect sites in LiCoO$_2$, 4 in LiFePO$_4$ and 3 in LiAlO$_2$. The structures for these defects were obtained by enumerating all the possible symmetrically unique combination of defects using the python library \verb|bsym|. Swapping of defects in these materials caused very small changes in the total energy/atom: the energies vary within a range of 40 meV/atom. In all three cases, both methods were able to reflect the small variance in energies, with DIEP performing better than M3GNET in 2 of the 3 cases.

Next, we consider H interstitial defects. An extra H is inserted into a crystal void of metallic structures and perovskites using the \verb|add_interstitial()| function from the \verb|OgStructure| class in the \verb|oganesson| python package. The presence of interstitial H in metals is one of the key causes of their embrittlement,\cite{lynch_hydrogen_2011} and occurs during the diffusion of hydrogen in perovksites en route of water splitting. Predicting the energy of a single interstitial atom in both metals and perovskites was challenging for DIEP and M3GNET. \textcolor{black}{For metals, the highest errors were due to H interstitial defects in crystal with triclinic symmetry (MAE for DIEP is 110 meV/atom, for M3GNET is 160 meV/atom). While DIEP is more accurate than M3GNET in predicting the energies of cubic and triclinic crystals, M3GNET surpassed the accuracy of DIEP for monoclinic (DIEP: 86 meV/atom, M3GNET: 80 meV/atom), orthorhombic (DIEP: 99 meV/atom, M3GNET: 92 meV/atom), trigonal (DIEP: 105 meV/atoms, M3GNET 73 meV/atom) and tetragonal (DIEP: 101 meV/atom, M3GNET: 86 meV/atoms) crystals.} Adding more interstitial H atoms in a selection of metals (Fe, Pt, Ni and Zn) improved the accuracy of the DIEP model, as shown in the bottom of Table \ref{tab_results_deformations}.

\begin{figure}[H]
	\begin{center}	
\includegraphics[width=0.5\textwidth]{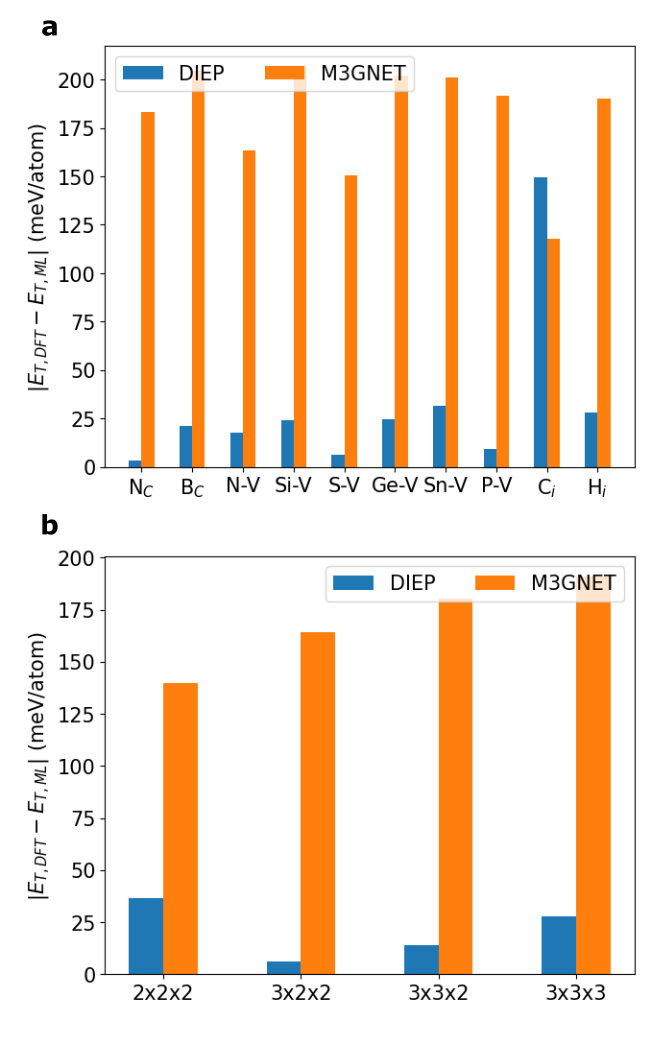}
\end{center}

\caption{\label{fig_results_defects} Prediction results for deformations. (a) The MAE for predicting the total energy/atom for each of the 10 diamond defects. (b) An examination of the influence of defect dilution on the quality of the prediction.}
\end{figure}

\noindent\textbf{Diamond defects} We test the predictions of DIEP and M3GNET for the following neutral diamond defects against the DFT values: nitrogen vacancy (N-$V$),\cite{rondin_magnetometry_2014} oxygen vacancy (O-$V$),\cite{thiering_characterization_2016} phosphorus vacancy (P-$V$), germanium vacancy (Ge-$V$),\cite{iwasaki_germanium-vacancy_2015} tin vacancy (Sn-$V$),\cite{iwasaki_tin-vacancy_2017} silicon vacancy (Si-$V$),\cite{flatae_silicon-vacancy_2020} sulfur vacancy (S-$V$),\cite{baker_electron_2008} substitutional boron (B$_C$),\cite{mainwood_substitutional_1979} substitutional nitrogen (N$_C$),\cite{kalish_search_2001,mainwood_substitutional_1979} substitutional oxygen (O$_C$),\cite{thiering_characterization_2016} interstitial hydrogen (H$_i$),\cite{goss_theory_2002} and interstitial carbon (C$_i$).\cite{weigel_carbon_1973} Figure \ref{fig_results_defects}a shows that DIEP achieves better predictive accuracy for all of these defects than M3GNET, except for the C$_i$ defect where the two models are close. The most challenging defect for both models is C$_i$, owing to the presence of bond distances that lie outside of the distribution of bonds in TinyUnitCells. We further examine the impact of combining two defects within the supercell. We generate a permutation of P-$V$, P-$V$, S-$V$, Sn-$V$, Si-$V$, N-$V$, O-$V$ with B$_C$, N$_C$ and O$_C$ (a total of 18 two-site defects). The accuracy of DIEP in this set surpasses that of M3GNET with an MAE of 22 meV versus 196 meV, respectively.

Further, we examine the impact of dilution on the quality of the predictions. Defect dilution means that a single defect is created in increasingly large supercells. Here we focus on the single carbon vacancy defect. Figure \ref{fig_results_defects}b displays the prediction error for the defect, starting with the defect in a $2\times 2\times 2$ diamond supercell, up to $3\times 3\times 3$. The MAE of the prediction of both M3GNET and DIEP reduces as the defect becomes more dilute, with DIEP highly surpassing the accuracy of M3GNET at all defect concentrations considered.

\subsection{Prediction of the potential energy surface} 


Training an ML model to predict the potential energy surface (PES)\cite{simons_walking_1983} requires the ability of the model to predict the potential energy as well as the atomic force vectors. The equivalence between the negative of the neural network's gradients with respect to the atomic coordinates and the forces on those atoms along those coordinates, is a natural connection between neural learning and the PES that was observed by Behler and Parrinello\cite{behler_generalized_2007} in their early work on neural network regression of material properties. To train the models, we use the MPF.2021.2.8 dataset which includes the total energy/atom, atomic forces and lattice stresses for nearly 188k structures. We applied the same network architecture as that used for M3GNET training. \textcolor{black}{For the network parameters $l_{max}$ and $n_{max}$, we test the following combination of values to find the combination $(l_{max},n_{max}$) that will yield the lowest validation error for the total energy/atom: (2, 1), (2, 2) and (3, 3).} We applied a training-validation-test partitioning of $90\%,5\%,5\%$ and only included structures with atomic forces within the range $-10$ and $10$ eV/\AA, following the procedure in Ref.~\citen{qi_robust_2024}. We label the PES trained model DIEP-PES, and compare its performance against the M3GNET-PES model released as part of the \verb|matgl| python library (\verb|M3GNet-MP-2021.2.8-PES|).\cite{chen_universal_2022} We train the DIEP-PES model by optimising the total energy per atom and the atomic forces, without optimising the lattice stress. \textcolor{black}{The combination $(l_{max},n_{max}$ that gave the lowest validation MAE for the total energy/atom was (3, 3), where the validation set error is 41 meV/atom.} For the test set, the DIEP-PES models achieves a total energy per atom MAE of 61 meV/atom and a force MAE of 73 meV/\AA. The total energy per atom MAE for DIEP-PES is higher than that of the M3GNET-PES (which is 34 meV/atom) whereas the force MAE of the DIEP-PES is close to that of the M3GNET-PES (70 meV/\AA). 

To assess the computation cost for running a PES calculation, we performed a single-point calculation with both DIEP-PES and M3GNET-PES on each structure of the entire MP database (155k structures). The average execution time of DIEP-PES on 1 GPU processor is 5.3 milliseconds/atom, while that of M3GNET-PES is 5.5 milliseconds/atom. We further examined the computational cost comparison by performing single-point calculations with DIEP-PES, M3GNET-PES and DFT (using VASP) for a random selection of 96 structures from the MP database where the number of atoms exceeds 30 atoms. The mean number of atoms in the set is 59 atoms. The DIEP-PES on 1 GPU processor consumed 1.4 milliseconds/atom, M3GNET-PES 1.3 milliseconds/atom, while the DFT calculation on 96 CPU cores consumed 18.1 seconds/atom. \textcolor{black}{This shows that the computational complexities of the two methods are very close, and that the computational performance of the ML methods exceeds that of DFT by 3 orders of magnitude.}

\begin{figure}[H]
	\begin{center}	
		\includegraphics[width=0.5\textwidth]{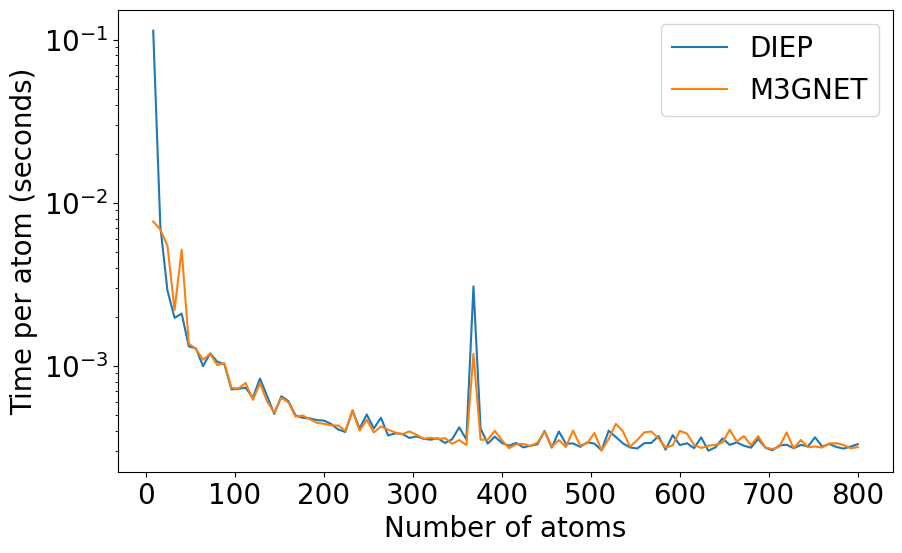}
	\end{center}
	
	\caption{\label{fig_computational_performance} The scaling behaviour of DIEP-PES and M3GNET-PES: the computational time per atom as the size of the system increases.}
\end{figure}

\textcolor{black}{The computational performance per atom improves with larger number of atoms. To see how the structure size influences the computational performance, we perform a single-point calculation on a supercell of diamond with size $N\times 1\times 1$ where $N$ takes values from 1 to 100. The number of atoms at $N=1$ is 8. The result is displayed in Figure \ref{fig_computational_performance}, showing that the computational scaling behaviours of both DIEP-PES and M3GNET-PES are very close.}

\subsubsection{Structure discovery using genetic algorithms} 

We examine the ability of DIEP-PES to re-discover the stable crystal structure for a number of binary compounds, by employing the genetic optimisation library available in the \verb*|oganesson| python package which wraps the genetic library in \verb*|ase|. We compare the number of compounds that we successfully re-discovered by DIEP-PES against those re-discovered by M3GNET. We display the results in Table \ref{tab_ga}.

We examine the ability of DIEP-PES to re-discover the stable crystal structure for a number of binary compounds with an energy above hull of zero eV/atom. We utilise the \verb|oganesson| \verb|GA| class which is a wrapper for the genetic algorithm code within the \verb|ase| python library. We limit the genetic optimisation process to an initial population of 20 structures, and 5 evolutions where 20 candidate structures are generated every evolution. \textcolor{black}{The genetic algorithm uses the total DFT energy as the fitness function. The following genetic evolution operators are allowed: cut and splice pairing,\cite{vilhelmsen_systematic_2012} soft mutation\cite{lyakhov_how_2010} and strain mutation,\cite{glass_uspexevolutionary_2006} with probabilities of 0.4, 0.3 and 0.3, respectively.} To compare between the optimised structures, we calculate the root mean square distance between two structures using the \verb|StructureMatcher.get_rms_dist()| method in the \verb|pymatgen| python library. The results are displayed in Table \ref{tab_ga}. The DIEP-PES was able to score more hits than M3GNET-PES: the DIEP-PES optimiser re-discovered the equilibrium structure of 12 out of 22 structures, whereas M3GNET-PES re-discovered 10 structures.

\begin{table}[t]
	\caption{The root mean square distance between the original structure and the structure obtained from genetic optimisation using DIEP and M3GNET models as PES calculators.}
	\label{tab_ga}
	\begin{center}
			\begin{tabular}{|ll|l|l|}
				
\hline
\textbf{MP ID}	&	\textbf{Formula}	&	\textbf{DIEP-PES}	&	\textbf{M3GNET-PES}	\\
\hline
mp-24208	&	CrH$_2$	&		&		\\
mp-784631	&	CrNi$_2$	&		&		\\
mp-182	&	SrGa$_2$	&	 0.01 	&		\\
mp-24728	&	VH$_2$	&		&		\\
mp-2732	&	PRh$_2$	&		&		\\
mp-11237	&	ScAg	&	 0.00 	&	 0.04 	\\
mp-1018138	&	VI$_2$	&		&		\\
mp-1000	&	BaTe	&	 0.00 	&	 0.00 	\\
mp-1169	&	ScCu	&	 0.00 	&	 0.01 	\\
mp-2516	&	YZn	&	 0.00 	&	 0.00 	\\
mp-1441	&	CsO$_2$	&		&		\\
mp-1883	&	SnTe	&	 0.00 	&		\\
mp-1008626	&	VTe$_2$	&		&		\\
mp-2221	&	Zr$_2$Ag	&		&		\\
mp-2697	&	SrO$_2$	&		&	 0.01 	\\
mp-2857	&	ScN	&	 0.02 	&	 0.01 	\\
mp-23251	&	KBr	&	 0.00 	&	 0.00 	\\
mp-987	&	ZnCu	&	 0.00 	&	 0.00 	\\
mp-2658	&	AlFe	&	 0.00 	&	 0.00 	\\
mp-28013	&	MnI$_2$	&	 0.29 	&		\\
mp-1207380	&	ZrIn	&	 0.00 	&	 0.00 	\\

\hline
\end{tabular}
\end{center}
\end{table}

\subsubsection{Carbon nanotube fracture} 
\begin{figure}[t]
	\begin{center}	
		\includegraphics[width=0.5\textwidth]{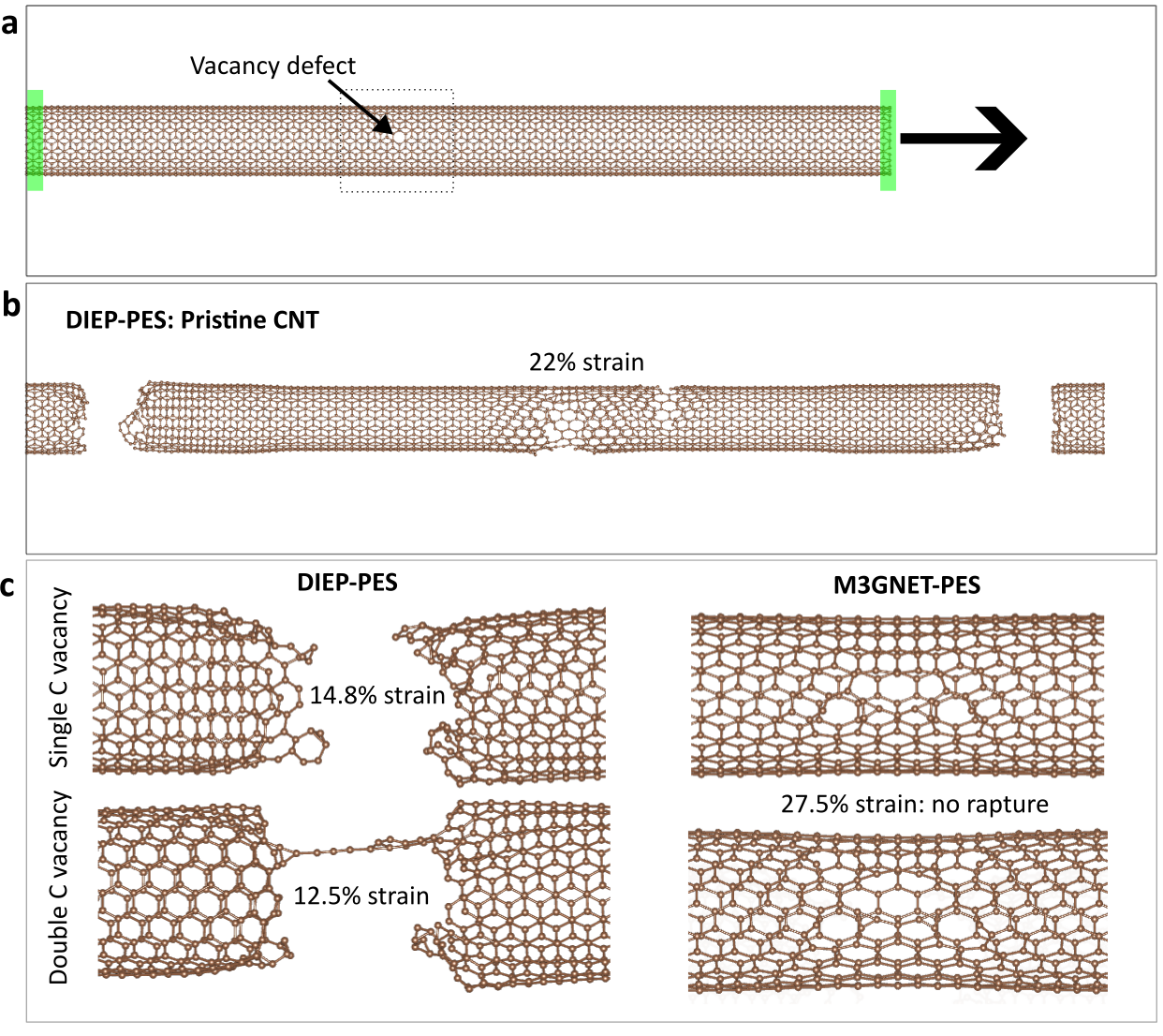}
	\end{center}
	
	\caption{\label{fig_cnt_fracture} \textbf{Simulation of CNT fracture.} (a) The simulation cell for the (10,10) zigzag CNT, highlighting the frozen atoms using green rectangles and indicating the carbon vacancy with an arrow. (b) For pristine CNT, it ruptures at a strain of \textcolor{black}{22\%} using the DIEP-PES, but never ruptures even when strain reaches 25\% using the M3GNET-PES. (c) The defective CNT ruptures at 14.8\% strain (single C vacancy) and at 12.5\% strain (double C vacancy) when the DIEP-PES is used. However, using the M3GNET-PES, the same defective CNTs never rupture, even when strain reaches 27.5\%. }
\end{figure}

A pristine carbon nanotube (CNT) can withstand an elongation strain up to 20\% of its original length, but will rupture at lower strains due to the presence of defects.\cite{yang_toughness_2016} We investigate the maximum strain that pristine and defective CNTs can withstand by performing quasi-static pulling of the CNT, in which a strain increment of 0.5\% is enforced on the structure followed by geometry optimisation until the CNT ruptures. This approach is similar to that in Refs. \citen{troya_carbon_2003,zhao_ultimate_2002,mielke_role_2004}. The CNT structure we simulate is a 172 \AA-long (10,10) zigzag CNT (2798 atoms), and we examine two defects: a single C vacancy defect, in which a single C atom close to the middle of the structure (as indicated in Figure. \ref{fig_cnt_fracture}a) is removed, and a double C vacancy defect, in which 2 C atoms close to the middle of the structure are removed. Following the removal of the C atoms, the atomic structure of the defect is optimised. For the pristine CNT, the structure ruptures when using DIEP-PIES at 22\% (Figure \ref{fig_cnt_fracture}b), which is close to the failure strain value of 20\% obtained using DFT for a (10,0) CNT.\cite{mielke_role_2004} However, the M3GNET-PES simulation does not lead to rupture. For vacancy defects, rupture is only produced when using the DIEP-PES: the CNT with a single C vacancy defect ruptures at a strain of 14.8\%, while that with a double C vacancy ruptures at 12.5\%, as displayed in Figure \ref{fig_cnt_fracture}c. The drop in failure strain from pristine to double vacancy defective structures qualitatively agree with the results in Ref.~\citen{yang_toughness_2016}, where a CNT with similar length was examined using classical molecular dynamics. We display the structure of the fractured CNT in Figure. \ref{fig_cnt_fracture}. \textcolor{black}{We also examined the influence of the distance between the C vacancies and the rupture strain by applying the strain procedure on two CNT structures where the vacancy-vacancy distances are 7.4~\AA~and 19.7~\\A. We found that the rupture strains are 15.4\% and 15.6\% respectively, indicating a slight decrease in CNT strength when the distance between the vacancies becomes smaller. The distance between the vacancies in the double vacancy structure in Figure \ref{fig_cnt_fracture} is the smallest possible, and hence is the weakest CNT structure (12.5\% rupture strain).} However, for M3GNET-PES, the structure does not rupture at all for an elongation strain of 27.5\%. That is, the M3GNET-PES highly overestimates the elasticity of the pristine and defective CNT structure, which is contrary to observation and theoretical calculations.

\section*{Conclusion} 

The ability to accurately predict the total energy is critical for the determination of the thermodynamic properties of materials. Our work improved the accuracy of predicting this quantity for materials with and without defects by introducing the direct integration of the external potential (DIEP) method. In principle, DIEP partially reflects the computations that take place during the  electronic structure optimisation step in density functional theory, and therefore more faithfully introduces ``physics'' in a physics-informed machine learning process. We demonstrated the enhanced accuracy of DIEP in predicting the total energy/atom for several datasets: a 12k dataset of pristine materials, and datasets that represent 6 classes of material imperfections. In addition, we established the ability of DIEP in predicting the potential energy surface for materials (total energies and forces) by performing structure optimisation and molecular dynamics tasks, in particular reproducing the maximum strain of a carbon nanotube structure. Enriched with its unique physical insight, DIEP is therefore suited for high-throughput screening procedures for accelerating material discovery.



\section*{Methods}

\noindent\textbf{DFT calculations} DFT calculations are performed using VASP 5.4.4.\cite{kresse_efficient_1996} The generalized gradient approximation (GGA) of Perdew, Burke and Ernzerfof (PBE),\cite{perdew_generalized_1996} The energy cut-off for the plane wave basis set is 520 eV, and the energy tolerance is $10^{-6}$ eV to ensure the accuracy of the calculations. In the structural energy minimization, the internal coordinates are allowed to relax until all of the forces are less than 0.01 eV/{\AA}. \textcolor{black}{For magnetic structures, the initial magnetic moments are set using the default VASP values.}



\section*{Supplementary Information}

The data points for Table \ref{tab_results_deformations} are available in the Supplementary Information file. The structures in the sheets of the Excel file are available in the structure repository \href{https://github.com/sheriftawfikabbas/materialsalchemist}{https://github.com/sheriftawfikabbas/materialsalchemist}.

\section*{Code availability}
DIEP was implemented on the codebase of Matgl and is available here: \href{https://github.com/sheriftawfikabbas/diep}{https://github.com/sheriftawfikabbas/diep}. The code that was used to generate the deformed structures is available as part of the \verb|oganesson| package: \href{https://github.com/sheriftawfikabbas/oganesson}{https://github.com/sheriftawfikabbas/oganesson}.

\section*{Data availability}

The VASP POSCAR files for the materials in TinyUnitCells and all of the defects examined in the work, the trained DIEP and M3GNET models are available here: \href{https://github.com/sheriftawfikabbas/materialsalchemist}{https://github.com/sheriftawfikabbas/materialsalchemist}. 

\bibliography{diip}

\begin{thebibliography}{38}
\expandafter\ifx\csname natexlab\endcsname\relax\def\natexlab#1{#1}\fi
\expandafter\ifx\csname bibnamefont\endcsname\relax
  \def\bibnamefont#1{#1}\fi
\expandafter\ifx\csname bibfnamefont\endcsname\relax
  \def\bibfnamefont#1{#1}\fi
\expandafter\ifx\csname citenamefont\endcsname\relax
  \def\citenamefont#1{#1}\fi
\expandafter\ifx\csname url\endcsname\relax
  \def\url#1{\texttt{#1}}\fi
\expandafter\ifx\csname urlprefix\endcsname\relax\def\urlprefix{URL }\fi
\providecommand{\bibinfo}[2]{#2}
\providecommand{\eprint}[2][]{\url{#2}}

\bibitem[{\citenamefont{Baskin et~al.}(1997)\citenamefont{Baskin, Palyulin, and
  Zefirov}}]{baskin_neural_1997}
\bibinfo{author}{\bibfnamefont{I.~I.} \bibnamefont{Baskin}},
  \bibinfo{author}{\bibfnamefont{V.~A.} \bibnamefont{Palyulin}},
  \bibnamefont{and} \bibinfo{author}{\bibfnamefont{N.~S.}
  \bibnamefont{Zefirov}}, \bibinfo{journal}{J. Chem. Inf. Comput. Sci.}
  \textbf{\bibinfo{volume}{37}}, \bibinfo{pages}{715} (\bibinfo{year}{1997}),
  ISSN \bibinfo{issn}{0095-2338},
  \urlprefix\url{https://pubs.acs.org/doi/10.1021/ci940128y}.

\bibitem[{\citenamefont{Zhang et~al.}(2023)\citenamefont{Zhang, Wang, Helwig,
  Luo, Fu, Xie, Liu, Lin, Xu, Yan et~al.}}]{zhang_artificial_nodate}
\bibinfo{author}{\bibfnamefont{X.}~\bibnamefont{Zhang}},
  \bibinfo{author}{\bibfnamefont{L.}~\bibnamefont{Wang}},
  \bibinfo{author}{\bibfnamefont{J.}~\bibnamefont{Helwig}},
  \bibinfo{author}{\bibfnamefont{Y.}~\bibnamefont{Luo}},
  \bibinfo{author}{\bibfnamefont{C.}~\bibnamefont{Fu}},
  \bibinfo{author}{\bibfnamefont{Y.}~\bibnamefont{Xie}},
  \bibinfo{author}{\bibfnamefont{M.}~\bibnamefont{Liu}},
  \bibinfo{author}{\bibfnamefont{Y.}~\bibnamefont{Lin}},
  \bibinfo{author}{\bibfnamefont{Z.}~\bibnamefont{Xu}},
  \bibinfo{author}{\bibfnamefont{K.}~\bibnamefont{Yan}}, \bibnamefont{et~al.}
  (\bibinfo{year}{2023}),
  \urlprefix\url{https://doi.org/10.48550/arXiv.2307.08423}.

\bibitem[{\citenamefont{Jain et~al.}(2013)\citenamefont{Jain, Ong, Hautier,
  Chen, Richards, Dacek, Cholia, Gunter, Skinner, Ceder
  et~al.}}]{jain_commentary_2013}
\bibinfo{author}{\bibfnamefont{A.}~\bibnamefont{Jain}},
  \bibinfo{author}{\bibfnamefont{S.~P.} \bibnamefont{Ong}},
  \bibinfo{author}{\bibfnamefont{G.}~\bibnamefont{Hautier}},
  \bibinfo{author}{\bibfnamefont{W.}~\bibnamefont{Chen}},
  \bibinfo{author}{\bibfnamefont{W.~D.} \bibnamefont{Richards}},
  \bibinfo{author}{\bibfnamefont{S.}~\bibnamefont{Dacek}},
  \bibinfo{author}{\bibfnamefont{S.}~\bibnamefont{Cholia}},
  \bibinfo{author}{\bibfnamefont{D.}~\bibnamefont{Gunter}},
  \bibinfo{author}{\bibfnamefont{D.}~\bibnamefont{Skinner}},
  \bibinfo{author}{\bibfnamefont{G.}~\bibnamefont{Ceder}},
  \bibnamefont{et~al.}, \bibinfo{journal}{APL Materials}
  \textbf{\bibinfo{volume}{1}}, \bibinfo{pages}{011002} (\bibinfo{year}{2013}),
  ISSN \bibinfo{issn}{2166-532X},
  \urlprefix\url{http://aip.scitation.org/doi/10.1063/1.4812323}.

\bibitem[{\citenamefont{Choudhary et~al.}(2020)\citenamefont{Choudhary,
  Garrity, Reid, DeCost, Biacchi, Hight~Walker, Trautt, Hattrick-Simpers,
  Kusne, Centrone et~al.}}]{choudhary_joint_2020}
\bibinfo{author}{\bibfnamefont{K.}~\bibnamefont{Choudhary}},
  \bibinfo{author}{\bibfnamefont{K.~F.} \bibnamefont{Garrity}},
  \bibinfo{author}{\bibfnamefont{A.~C.~E.} \bibnamefont{Reid}},
  \bibinfo{author}{\bibfnamefont{B.}~\bibnamefont{DeCost}},
  \bibinfo{author}{\bibfnamefont{A.~J.} \bibnamefont{Biacchi}},
  \bibinfo{author}{\bibfnamefont{A.~R.} \bibnamefont{Hight~Walker}},
  \bibinfo{author}{\bibfnamefont{Z.}~\bibnamefont{Trautt}},
  \bibinfo{author}{\bibfnamefont{J.}~\bibnamefont{Hattrick-Simpers}},
  \bibinfo{author}{\bibfnamefont{A.~G.} \bibnamefont{Kusne}},
  \bibinfo{author}{\bibfnamefont{A.}~\bibnamefont{Centrone}},
  \bibnamefont{et~al.}, \bibinfo{journal}{npj Comput Mater}
  \textbf{\bibinfo{volume}{6}}, \bibinfo{pages}{173} (\bibinfo{year}{2020}),
  ISSN \bibinfo{issn}{2057-3960},
  \urlprefix\url{https://www.nature.com/articles/s41524-020-00440-1}.

\bibitem[{\citenamefont{Curtarolo et~al.}(2012)\citenamefont{Curtarolo,
  Setyawan, Hart, Jahnatek, Chepulskii, Taylor, Wang, Xue, Yang, Levy
  et~al.}}]{curtarolo_aflow_2012}
\bibinfo{author}{\bibfnamefont{S.}~\bibnamefont{Curtarolo}},
  \bibinfo{author}{\bibfnamefont{W.}~\bibnamefont{Setyawan}},
  \bibinfo{author}{\bibfnamefont{G.~L.} \bibnamefont{Hart}},
  \bibinfo{author}{\bibfnamefont{M.}~\bibnamefont{Jahnatek}},
  \bibinfo{author}{\bibfnamefont{R.~V.} \bibnamefont{Chepulskii}},
  \bibinfo{author}{\bibfnamefont{R.~H.} \bibnamefont{Taylor}},
  \bibinfo{author}{\bibfnamefont{S.}~\bibnamefont{Wang}},
  \bibinfo{author}{\bibfnamefont{J.}~\bibnamefont{Xue}},
  \bibinfo{author}{\bibfnamefont{K.}~\bibnamefont{Yang}},
  \bibinfo{author}{\bibfnamefont{O.}~\bibnamefont{Levy}}, \bibnamefont{et~al.},
  \bibinfo{journal}{Computational Materials Science}
  \textbf{\bibinfo{volume}{58}}, \bibinfo{pages}{218} (\bibinfo{year}{2012}),
  ISSN \bibinfo{issn}{09270256},
  \urlprefix\url{https://linkinghub.elsevier.com/retrieve/pii/S0927025612000717}.

\bibitem[{\citenamefont{Kirklin et~al.}(2015)\citenamefont{Kirklin, Saal,
  Meredig, Thompson, Doak, Aykol, Rühl, and Wolverton}}]{kirklin_open_2015}
\bibinfo{author}{\bibfnamefont{S.}~\bibnamefont{Kirklin}},
  \bibinfo{author}{\bibfnamefont{J.~E.} \bibnamefont{Saal}},
  \bibinfo{author}{\bibfnamefont{B.}~\bibnamefont{Meredig}},
  \bibinfo{author}{\bibfnamefont{A.}~\bibnamefont{Thompson}},
  \bibinfo{author}{\bibfnamefont{J.~W.} \bibnamefont{Doak}},
  \bibinfo{author}{\bibfnamefont{M.}~\bibnamefont{Aykol}},
  \bibinfo{author}{\bibfnamefont{S.}~\bibnamefont{Rühl}}, \bibnamefont{and}
  \bibinfo{author}{\bibfnamefont{C.}~\bibnamefont{Wolverton}},
  \bibinfo{journal}{npj Comput Mater} \textbf{\bibinfo{volume}{1}},
  \bibinfo{pages}{15010} (\bibinfo{year}{2015}), ISSN
  \bibinfo{issn}{2057-3960},
  \urlprefix\url{http://www.nature.com/articles/npjcompumats201510}.

\bibitem[{\citenamefont{Gasteiger et~al.}(2022)\citenamefont{Gasteiger, Groß,
  and Günnemann}}]{gasteiger_directional_2022}
\bibinfo{author}{\bibfnamefont{J.}~\bibnamefont{Gasteiger}},
  \bibinfo{author}{\bibfnamefont{J.}~\bibnamefont{Groß}}, \bibnamefont{and}
  \bibinfo{author}{\bibfnamefont{S.}~\bibnamefont{Günnemann}},
  \emph{\bibinfo{title}{Directional {Message} {Passing} for {Molecular}
  {Graphs}}} (\bibinfo{year}{2022}), \bibinfo{note}{arXiv:2003.03123 [physics,
  stat]}, \urlprefix\url{http://arxiv.org/abs/2003.03123}.

\bibitem[{\citenamefont{Chen and Ong}(2022)}]{chen_universal_2022}
\bibinfo{author}{\bibfnamefont{C.}~\bibnamefont{Chen}} \bibnamefont{and}
  \bibinfo{author}{\bibfnamefont{S.~P.} \bibnamefont{Ong}},
  \bibinfo{journal}{Nat Comput Sci} \textbf{\bibinfo{volume}{2}},
  \bibinfo{pages}{718} (\bibinfo{year}{2022}), ISSN \bibinfo{issn}{2662-8457},
  \urlprefix\url{https://www.nature.com/articles/s43588-022-00349-3}.

\bibitem[{\citenamefont{Batatia et~al.}(2023)\citenamefont{Batatia, Kovács,
  Simm, Ortner, and Csányi}}]{batatia_mace_2023}
\bibinfo{author}{\bibfnamefont{I.}~\bibnamefont{Batatia}},
  \bibinfo{author}{\bibfnamefont{D.~P.} \bibnamefont{Kovács}},
  \bibinfo{author}{\bibfnamefont{G.~N.~C.} \bibnamefont{Simm}},
  \bibinfo{author}{\bibfnamefont{C.}~\bibnamefont{Ortner}}, \bibnamefont{and}
  \bibinfo{author}{\bibfnamefont{G.}~\bibnamefont{Csányi}},
  \emph{\bibinfo{title}{{MACE}: {Higher} {Order} {Equivariant} {Message}
  {Passing} {Neural} {Networks} for {Fast} and {Accurate} {Force} {Fields}}}
  (\bibinfo{year}{2023}), \bibinfo{note}{arXiv:2206.07697 [cond-mat,
  physics:physics, stat]}, \urlprefix\url{http://arxiv.org/abs/2206.07697}.

\bibitem[{\citenamefont{Hohenberg and Kohn}(1964)}]{PhysRev.136.B864}
\bibinfo{author}{\bibfnamefont{P.}~\bibnamefont{Hohenberg}} \bibnamefont{and}
  \bibinfo{author}{\bibfnamefont{W.}~\bibnamefont{Kohn}},
  \bibinfo{journal}{Phys. Rev.} \textbf{\bibinfo{volume}{136}},
  \bibinfo{pages}{B864} (\bibinfo{year}{1964}),
  \urlprefix\url{https://link.aps.org/doi/10.1103/PhysRev.136.B864}.

\bibitem[{\citenamefont{Cuevas-Zuviría and
  Pacios}(2020)}]{cuevas-zuviria_analytical_2020}
\bibinfo{author}{\bibfnamefont{B.}~\bibnamefont{Cuevas-Zuviría}}
  \bibnamefont{and} \bibinfo{author}{\bibfnamefont{L.~F.}
  \bibnamefont{Pacios}}, \bibinfo{journal}{J. Chem. Inf. Model.}
  \textbf{\bibinfo{volume}{60}}, \bibinfo{pages}{3831} (\bibinfo{year}{2020}),
  ISSN \bibinfo{issn}{1549-9596, 1549-960X},
  \urlprefix\url{https://pubs.acs.org/doi/10.1021/acs.jcim.0c00197}.

\bibitem[{\citenamefont{Sunshine et~al.}(2023)\citenamefont{Sunshine, Shuaibi,
  Ulissi, and Kitchin}}]{sunshine_chemical_2023}
\bibinfo{author}{\bibfnamefont{E.~M.} \bibnamefont{Sunshine}},
  \bibinfo{author}{\bibfnamefont{M.}~\bibnamefont{Shuaibi}},
  \bibinfo{author}{\bibfnamefont{Z.~W.} \bibnamefont{Ulissi}},
  \bibnamefont{and} \bibinfo{author}{\bibfnamefont{J.~R.}
  \bibnamefont{Kitchin}}, \bibinfo{journal}{J. Phys. Chem. C}
  \textbf{\bibinfo{volume}{127}}, \bibinfo{pages}{23459}
  (\bibinfo{year}{2023}), ISSN \bibinfo{issn}{1932-7447, 1932-7455},
  \urlprefix\url{https://pubs.acs.org/doi/10.1021/acs.jpcc.3c06157}.

\bibitem[{\citenamefont{Gong et~al.}(2019)\citenamefont{Gong, Xie, Zhu, Wang,
  Fadel, Li, and Grossman}}]{gong_predicting_2019}
\bibinfo{author}{\bibfnamefont{S.}~\bibnamefont{Gong}},
  \bibinfo{author}{\bibfnamefont{T.}~\bibnamefont{Xie}},
  \bibinfo{author}{\bibfnamefont{T.}~\bibnamefont{Zhu}},
  \bibinfo{author}{\bibfnamefont{S.}~\bibnamefont{Wang}},
  \bibinfo{author}{\bibfnamefont{E.~R.} \bibnamefont{Fadel}},
  \bibinfo{author}{\bibfnamefont{Y.}~\bibnamefont{Li}}, \bibnamefont{and}
  \bibinfo{author}{\bibfnamefont{J.~C.} \bibnamefont{Grossman}},
  \bibinfo{journal}{Phys. Rev. B} \textbf{\bibinfo{volume}{100}},
  \bibinfo{pages}{184103} (\bibinfo{year}{2019}), ISSN
  \bibinfo{issn}{2469-9950, 2469-9969},
  \urlprefix\url{https://link.aps.org/doi/10.1103/PhysRevB.100.184103}.

\bibitem[{\citenamefont{Jørgensen and Bhowmik}(2020)}]{jorgensen_deepdft_2020}
\bibinfo{author}{\bibfnamefont{P.~B.} \bibnamefont{Jørgensen}}
  \bibnamefont{and} \bibinfo{author}{\bibfnamefont{A.}~\bibnamefont{Bhowmik}},
  \emph{\bibinfo{title}{{DeepDFT}: {Neural} {Message} {Passing} {Network} for
  {Accurate} {Charge} {Density} {Prediction}}} (\bibinfo{year}{2020}),
  \bibinfo{note}{arXiv:2011.03346 [physics]},
  \urlprefix\url{http://arxiv.org/abs/2011.03346}.

\bibitem[{\citenamefont{Batchelor et~al.}(2019)\citenamefont{Batchelor,
  Pedersen, Winther, Castelli, Jacobsen, and
  Rossmeisl}}]{batchelor_high-entropy_2019}
\bibinfo{author}{\bibfnamefont{T.~A.} \bibnamefont{Batchelor}},
  \bibinfo{author}{\bibfnamefont{J.~K.} \bibnamefont{Pedersen}},
  \bibinfo{author}{\bibfnamefont{S.~H.} \bibnamefont{Winther}},
  \bibinfo{author}{\bibfnamefont{I.~E.} \bibnamefont{Castelli}},
  \bibinfo{author}{\bibfnamefont{K.~W.} \bibnamefont{Jacobsen}},
  \bibnamefont{and}
  \bibinfo{author}{\bibfnamefont{J.}~\bibnamefont{Rossmeisl}},
  \bibinfo{journal}{Joule} \textbf{\bibinfo{volume}{3}}, \bibinfo{pages}{834}
  (\bibinfo{year}{2019}), ISSN \bibinfo{issn}{25424351},
  \urlprefix\url{https://linkinghub.elsevier.com/retrieve/pii/S2542435118306214}.

\bibitem[{\citenamefont{Lynch}(2011)}]{lynch_hydrogen_2011}
\bibinfo{author}{\bibfnamefont{S.}~\bibnamefont{Lynch}}, in
  \emph{\bibinfo{booktitle}{Stress {Corrosion} {Cracking}}}
  (\bibinfo{publisher}{Elsevier}, \bibinfo{year}{2011}), pp.
  \bibinfo{pages}{90--130}, ISBN \bibinfo{isbn}{978-1-84569-673-3},
  \urlprefix\url{https://linkinghub.elsevier.com/retrieve/pii/B978184569673350002X}.

\bibitem[{\citenamefont{Rondin et~al.}(2014)\citenamefont{Rondin, Tetienne,
  Hingant, Roch, Maletinsky, and Jacques}}]{rondin_magnetometry_2014}
\bibinfo{author}{\bibfnamefont{L.}~\bibnamefont{Rondin}},
  \bibinfo{author}{\bibfnamefont{J.-P.} \bibnamefont{Tetienne}},
  \bibinfo{author}{\bibfnamefont{T.}~\bibnamefont{Hingant}},
  \bibinfo{author}{\bibfnamefont{J.-F.} \bibnamefont{Roch}},
  \bibinfo{author}{\bibfnamefont{P.}~\bibnamefont{Maletinsky}},
  \bibnamefont{and} \bibinfo{author}{\bibfnamefont{V.}~\bibnamefont{Jacques}},
  \bibinfo{journal}{Reports on Progress in Physics}
  \textbf{\bibinfo{volume}{77}}, \bibinfo{pages}{056503}
  (\bibinfo{year}{2014}), ISSN \bibinfo{issn}{0034-4885, 1361-6633},
  \urlprefix\url{https://iopscience.iop.org/article/10.1088/0034-4885/77/5/056503}.

\bibitem[{\citenamefont{Thiering and
  Gali}(2016)}]{thiering_characterization_2016}
\bibinfo{author}{\bibfnamefont{G.}~\bibnamefont{Thiering}} \bibnamefont{and}
  \bibinfo{author}{\bibfnamefont{A.}~\bibnamefont{Gali}},
  \bibinfo{journal}{Physical Review B} \textbf{\bibinfo{volume}{94}},
  \bibinfo{pages}{125202} (\bibinfo{year}{2016}), ISSN
  \bibinfo{issn}{2469-9950, 2469-9969},
  \urlprefix\url{https://link.aps.org/doi/10.1103/PhysRevB.94.125202}.

\bibitem[{\citenamefont{Iwasaki et~al.}(2015)\citenamefont{Iwasaki, Ishibashi,
  Miyamoto, Doi, Kobayashi, Miyazaki, Tahara, Jahnke, Rogers, Naydenov
  et~al.}}]{iwasaki_germanium-vacancy_2015}
\bibinfo{author}{\bibfnamefont{T.}~\bibnamefont{Iwasaki}},
  \bibinfo{author}{\bibfnamefont{F.}~\bibnamefont{Ishibashi}},
  \bibinfo{author}{\bibfnamefont{Y.}~\bibnamefont{Miyamoto}},
  \bibinfo{author}{\bibfnamefont{Y.}~\bibnamefont{Doi}},
  \bibinfo{author}{\bibfnamefont{S.}~\bibnamefont{Kobayashi}},
  \bibinfo{author}{\bibfnamefont{T.}~\bibnamefont{Miyazaki}},
  \bibinfo{author}{\bibfnamefont{K.}~\bibnamefont{Tahara}},
  \bibinfo{author}{\bibfnamefont{K.~D.} \bibnamefont{Jahnke}},
  \bibinfo{author}{\bibfnamefont{L.~J.} \bibnamefont{Rogers}},
  \bibinfo{author}{\bibfnamefont{B.}~\bibnamefont{Naydenov}},
  \bibnamefont{et~al.}, \bibinfo{journal}{Scientific Reports}
  \textbf{\bibinfo{volume}{5}}, \bibinfo{pages}{12882} (\bibinfo{year}{2015}),
  ISSN \bibinfo{issn}{2045-2322},
  \urlprefix\url{https://www.nature.com/articles/srep12882}.

\bibitem[{\citenamefont{Iwasaki et~al.}(2017)\citenamefont{Iwasaki, Miyamoto,
  Taniguchi, Siyushev, Metsch, Jelezko, and Hatano}}]{iwasaki_tin-vacancy_2017}
\bibinfo{author}{\bibfnamefont{T.}~\bibnamefont{Iwasaki}},
  \bibinfo{author}{\bibfnamefont{Y.}~\bibnamefont{Miyamoto}},
  \bibinfo{author}{\bibfnamefont{T.}~\bibnamefont{Taniguchi}},
  \bibinfo{author}{\bibfnamefont{P.}~\bibnamefont{Siyushev}},
  \bibinfo{author}{\bibfnamefont{M.~H.} \bibnamefont{Metsch}},
  \bibinfo{author}{\bibfnamefont{F.}~\bibnamefont{Jelezko}}, \bibnamefont{and}
  \bibinfo{author}{\bibfnamefont{M.}~\bibnamefont{Hatano}},
  \bibinfo{journal}{Physical Review Letters} \textbf{\bibinfo{volume}{119}},
  \bibinfo{pages}{253601} (\bibinfo{year}{2017}), ISSN
  \bibinfo{issn}{0031-9007, 1079-7114},
  \urlprefix\url{https://link.aps.org/doi/10.1103/PhysRevLett.119.253601}.

\bibitem[{\citenamefont{Flatae et~al.}(2020)\citenamefont{Flatae, Lagomarsino,
  Sledz, Soltani, Nicley, Haenen, Rechenberg, Becker, Sciortino, Gelli
  et~al.}}]{flatae_silicon-vacancy_2020}
\bibinfo{author}{\bibfnamefont{A.~M.} \bibnamefont{Flatae}},
  \bibinfo{author}{\bibfnamefont{S.}~\bibnamefont{Lagomarsino}},
  \bibinfo{author}{\bibfnamefont{F.}~\bibnamefont{Sledz}},
  \bibinfo{author}{\bibfnamefont{N.}~\bibnamefont{Soltani}},
  \bibinfo{author}{\bibfnamefont{S.~S.} \bibnamefont{Nicley}},
  \bibinfo{author}{\bibfnamefont{K.}~\bibnamefont{Haenen}},
  \bibinfo{author}{\bibfnamefont{R.}~\bibnamefont{Rechenberg}},
  \bibinfo{author}{\bibfnamefont{M.~F.} \bibnamefont{Becker}},
  \bibinfo{author}{\bibfnamefont{S.}~\bibnamefont{Sciortino}},
  \bibinfo{author}{\bibfnamefont{N.}~\bibnamefont{Gelli}},
  \bibnamefont{et~al.}, \bibinfo{journal}{Diamond and Related Materials}
  \textbf{\bibinfo{volume}{105}}, \bibinfo{pages}{107797}
  (\bibinfo{year}{2020}), ISSN \bibinfo{issn}{09259635},
  \urlprefix\url{https://linkinghub.elsevier.com/retrieve/pii/S0925963519309744}.

\bibitem[{\citenamefont{Baker et~al.}(2008)\citenamefont{Baker, Van~Wyk, Goss,
  and Briddon}}]{baker_electron_2008}
\bibinfo{author}{\bibfnamefont{J.~M.} \bibnamefont{Baker}},
  \bibinfo{author}{\bibfnamefont{J.~A.} \bibnamefont{Van~Wyk}},
  \bibinfo{author}{\bibfnamefont{J.~P.} \bibnamefont{Goss}}, \bibnamefont{and}
  \bibinfo{author}{\bibfnamefont{P.~R.} \bibnamefont{Briddon}},
  \bibinfo{journal}{Physical Review B} \textbf{\bibinfo{volume}{78}},
  \bibinfo{pages}{235203} (\bibinfo{year}{2008}), ISSN
  \bibinfo{issn}{1098-0121, 1550-235X},
  \urlprefix\url{https://link.aps.org/doi/10.1103/PhysRevB.78.235203}.

\bibitem[{\citenamefont{Mainwood}(1979)}]{mainwood_substitutional_1979}
\bibinfo{author}{\bibfnamefont{A.}~\bibnamefont{Mainwood}},
  \bibinfo{journal}{Journal of Physics C: Solid State Physics}
  \textbf{\bibinfo{volume}{12}}, \bibinfo{pages}{2543} (\bibinfo{year}{1979}),
  ISSN \bibinfo{issn}{0022-3719},
  \urlprefix\url{https://iopscience.iop.org/article/10.1088/0022-3719/12/13/018}.

\bibitem[{\citenamefont{Kalish}(2001)}]{kalish_search_2001}
\bibinfo{author}{\bibfnamefont{R.}~\bibnamefont{Kalish}},
  \bibinfo{journal}{Diamond and Related Materials}
  \textbf{\bibinfo{volume}{10}}, \bibinfo{pages}{1749} (\bibinfo{year}{2001}),
  ISSN \bibinfo{issn}{09259635},
  \urlprefix\url{https://linkinghub.elsevier.com/retrieve/pii/S0925963501004265}.

\bibitem[{\citenamefont{Goss et~al.}(2002)\citenamefont{Goss, Jones, Heggie,
  Ewels, Briddon, and Öberg}}]{goss_theory_2002}
\bibinfo{author}{\bibfnamefont{J.~P.} \bibnamefont{Goss}},
  \bibinfo{author}{\bibfnamefont{R.}~\bibnamefont{Jones}},
  \bibinfo{author}{\bibfnamefont{M.~I.} \bibnamefont{Heggie}},
  \bibinfo{author}{\bibfnamefont{C.~P.} \bibnamefont{Ewels}},
  \bibinfo{author}{\bibfnamefont{P.~R.} \bibnamefont{Briddon}},
  \bibnamefont{and} \bibinfo{author}{\bibfnamefont{S.}~\bibnamefont{Öberg}},
  \bibinfo{journal}{Physical Review B} \textbf{\bibinfo{volume}{65}},
  \bibinfo{pages}{115207} (\bibinfo{year}{2002}), ISSN
  \bibinfo{issn}{0163-1829, 1095-3795},
  \urlprefix\url{https://link.aps.org/doi/10.1103/PhysRevB.65.115207}.

\bibitem[{\citenamefont{Weigel et~al.}(1973)\citenamefont{Weigel, Peak,
  Corbett, Watkins, and Messmer}}]{weigel_carbon_1973}
\bibinfo{author}{\bibfnamefont{C.}~\bibnamefont{Weigel}},
  \bibinfo{author}{\bibfnamefont{D.}~\bibnamefont{Peak}},
  \bibinfo{author}{\bibfnamefont{J.~W.} \bibnamefont{Corbett}},
  \bibinfo{author}{\bibfnamefont{G.~D.} \bibnamefont{Watkins}},
  \bibnamefont{and} \bibinfo{author}{\bibfnamefont{R.~P.}
  \bibnamefont{Messmer}}, \bibinfo{journal}{Physical Review B}
  \textbf{\bibinfo{volume}{8}}, \bibinfo{pages}{2906} (\bibinfo{year}{1973}),
  ISSN \bibinfo{issn}{0556-2805},
  \urlprefix\url{https://link.aps.org/doi/10.1103/PhysRevB.8.2906}.

\bibitem[{\citenamefont{Simons et~al.}(1983)\citenamefont{Simons, Joergensen,
  Taylor, and Ozment}}]{simons_walking_1983}
\bibinfo{author}{\bibfnamefont{J.}~\bibnamefont{Simons}},
  \bibinfo{author}{\bibfnamefont{P.}~\bibnamefont{Joergensen}},
  \bibinfo{author}{\bibfnamefont{H.}~\bibnamefont{Taylor}}, \bibnamefont{and}
  \bibinfo{author}{\bibfnamefont{J.}~\bibnamefont{Ozment}},
  \bibinfo{journal}{J. Phys. Chem.} \textbf{\bibinfo{volume}{87}},
  \bibinfo{pages}{2745} (\bibinfo{year}{1983}), ISSN \bibinfo{issn}{0022-3654,
  1541-5740}, \urlprefix\url{https://pubs.acs.org/doi/abs/10.1021/j100238a013}.

\bibitem[{\citenamefont{Behler and Parrinello}(2007)}]{behler_generalized_2007}
\bibinfo{author}{\bibfnamefont{J.}~\bibnamefont{Behler}} \bibnamefont{and}
  \bibinfo{author}{\bibfnamefont{M.}~\bibnamefont{Parrinello}},
  \bibinfo{journal}{Phys. Rev. Lett.} \textbf{\bibinfo{volume}{98}},
  \bibinfo{pages}{146401} (\bibinfo{year}{2007}), ISSN
  \bibinfo{issn}{0031-9007, 1079-7114},
  \urlprefix\url{https://link.aps.org/doi/10.1103/PhysRevLett.98.146401}.

\bibitem[{\citenamefont{Qi et~al.}(2024)\citenamefont{Qi, Ko, Wood, Pham, and
  Ong}}]{qi_robust_2024}
\bibinfo{author}{\bibfnamefont{J.}~\bibnamefont{Qi}},
  \bibinfo{author}{\bibfnamefont{T.~W.} \bibnamefont{Ko}},
  \bibinfo{author}{\bibfnamefont{B.~C.} \bibnamefont{Wood}},
  \bibinfo{author}{\bibfnamefont{T.~A.} \bibnamefont{Pham}}, \bibnamefont{and}
  \bibinfo{author}{\bibfnamefont{S.~P.} \bibnamefont{Ong}},
  \bibinfo{journal}{npj Computational Materials} \textbf{\bibinfo{volume}{10}},
  \bibinfo{pages}{43} (\bibinfo{year}{2024}), ISSN \bibinfo{issn}{2057-3960},
  \urlprefix\url{https://www.nature.com/articles/s41524-024-01227-4}.

\bibitem[{\citenamefont{Vilhelmsen and
  Hammer}(2012)}]{vilhelmsen_systematic_2012}
\bibinfo{author}{\bibfnamefont{L.~B.} \bibnamefont{Vilhelmsen}}
  \bibnamefont{and} \bibinfo{author}{\bibfnamefont{B.}~\bibnamefont{Hammer}},
  \bibinfo{journal}{Phys. Rev. Lett.} \textbf{\bibinfo{volume}{108}},
  \bibinfo{pages}{126101} (\bibinfo{year}{2012}), ISSN
  \bibinfo{issn}{0031-9007, 1079-7114},
  \urlprefix\url{https://link.aps.org/doi/10.1103/PhysRevLett.108.126101}.

\bibitem[{\citenamefont{Lyakhov et~al.}(2010)\citenamefont{Lyakhov, Oganov, and
  Valle}}]{lyakhov_how_2010}
\bibinfo{author}{\bibfnamefont{A.~O.} \bibnamefont{Lyakhov}},
  \bibinfo{author}{\bibfnamefont{A.~R.} \bibnamefont{Oganov}},
  \bibnamefont{and} \bibinfo{author}{\bibfnamefont{M.}~\bibnamefont{Valle}},
  \bibinfo{journal}{Computer Physics Communications}
  \textbf{\bibinfo{volume}{181}}, \bibinfo{pages}{1623} (\bibinfo{year}{2010}),
  ISSN \bibinfo{issn}{00104655},
  \urlprefix\url{https://linkinghub.elsevier.com/retrieve/pii/S0010465510001840}.

\bibitem[{\citenamefont{Glass et~al.}(2006)\citenamefont{Glass, Oganov, and
  Hansen}}]{glass_uspexevolutionary_2006}
\bibinfo{author}{\bibfnamefont{C.~W.} \bibnamefont{Glass}},
  \bibinfo{author}{\bibfnamefont{A.~R.} \bibnamefont{Oganov}},
  \bibnamefont{and} \bibinfo{author}{\bibfnamefont{N.}~\bibnamefont{Hansen}},
  \bibinfo{journal}{Computer Physics Communications}
  \textbf{\bibinfo{volume}{175}}, \bibinfo{pages}{713} (\bibinfo{year}{2006}),
  ISSN \bibinfo{issn}{00104655},
  \urlprefix\url{https://linkinghub.elsevier.com/retrieve/pii/S0010465506002931}.

\bibitem[{\citenamefont{Yang et~al.}(2016)\citenamefont{Yang, Greenfeld, and
  Wagner}}]{yang_toughness_2016}
\bibinfo{author}{\bibfnamefont{L.}~\bibnamefont{Yang}},
  \bibinfo{author}{\bibfnamefont{I.}~\bibnamefont{Greenfeld}},
  \bibnamefont{and} \bibinfo{author}{\bibfnamefont{H.~D.}
  \bibnamefont{Wagner}}, \bibinfo{journal}{Science Advances}
  \textbf{\bibinfo{volume}{2}}, \bibinfo{pages}{e1500969}
  (\bibinfo{year}{2016}), ISSN \bibinfo{issn}{2375-2548},
  \urlprefix\url{https://www.science.org/doi/10.1126/sciadv.1500969}.

\bibitem[{\citenamefont{Troya et~al.}(2003)\citenamefont{Troya, Mielke, and
  Schatz}}]{troya_carbon_2003}
\bibinfo{author}{\bibfnamefont{D.}~\bibnamefont{Troya}},
  \bibinfo{author}{\bibfnamefont{S.~L.} \bibnamefont{Mielke}},
  \bibnamefont{and} \bibinfo{author}{\bibfnamefont{G.~C.}
  \bibnamefont{Schatz}}, \bibinfo{journal}{Chemical Physics Letters}
  \textbf{\bibinfo{volume}{382}}, \bibinfo{pages}{133} (\bibinfo{year}{2003}),
  ISSN \bibinfo{issn}{00092614},
  \urlprefix\url{https://linkinghub.elsevier.com/retrieve/pii/S000926140301830X}.

\bibitem[{\citenamefont{Zhao et~al.}(2002)\citenamefont{Zhao, Nardelli, and
  Bernholc}}]{zhao_ultimate_2002}
\bibinfo{author}{\bibfnamefont{Q.}~\bibnamefont{Zhao}},
  \bibinfo{author}{\bibfnamefont{M.~B.} \bibnamefont{Nardelli}},
  \bibnamefont{and} \bibinfo{author}{\bibfnamefont{J.}~\bibnamefont{Bernholc}},
  \bibinfo{journal}{Physical Review B} \textbf{\bibinfo{volume}{65}},
  \bibinfo{pages}{144105} (\bibinfo{year}{2002}), ISSN
  \bibinfo{issn}{0163-1829, 1095-3795},
  \urlprefix\url{https://link.aps.org/doi/10.1103/PhysRevB.65.144105}.

\bibitem[{\citenamefont{Mielke et~al.}(2004)\citenamefont{Mielke, Troya, Zhang,
  Li, Xiao, Car, Ruoff, Schatz, and Belytschko}}]{mielke_role_2004}
\bibinfo{author}{\bibfnamefont{S.~L.} \bibnamefont{Mielke}},
  \bibinfo{author}{\bibfnamefont{D.}~\bibnamefont{Troya}},
  \bibinfo{author}{\bibfnamefont{S.}~\bibnamefont{Zhang}},
  \bibinfo{author}{\bibfnamefont{J.-L.} \bibnamefont{Li}},
  \bibinfo{author}{\bibfnamefont{S.}~\bibnamefont{Xiao}},
  \bibinfo{author}{\bibfnamefont{R.}~\bibnamefont{Car}},
  \bibinfo{author}{\bibfnamefont{R.~S.} \bibnamefont{Ruoff}},
  \bibinfo{author}{\bibfnamefont{G.~C.} \bibnamefont{Schatz}},
  \bibnamefont{and}
  \bibinfo{author}{\bibfnamefont{T.}~\bibnamefont{Belytschko}},
  \bibinfo{journal}{Chemical Physics Letters} \textbf{\bibinfo{volume}{390}},
  \bibinfo{pages}{413} (\bibinfo{year}{2004}), ISSN \bibinfo{issn}{00092614},
  \urlprefix\url{https://linkinghub.elsevier.com/retrieve/pii/S0009261404005998}.

\bibitem[{\citenamefont{Kresse and Furthmüller}(1996)}]{kresse_efficient_1996}
\bibinfo{author}{\bibfnamefont{G.}~\bibnamefont{Kresse}} \bibnamefont{and}
  \bibinfo{author}{\bibfnamefont{J.}~\bibnamefont{Furthmüller}},
  \bibinfo{journal}{Physical Review B} \textbf{\bibinfo{volume}{54}},
  \bibinfo{pages}{11169} (\bibinfo{year}{1996}), ISSN \bibinfo{issn}{0163-1829,
  1095-3795},
  \urlprefix\url{https://link.aps.org/doi/10.1103/PhysRevB.54.11169}.

\bibitem[{\citenamefont{Perdew et~al.}(1996)\citenamefont{Perdew, Burke, and
  Ernzerhof}}]{perdew_generalized_1996}
\bibinfo{author}{\bibfnamefont{J.~P.} \bibnamefont{Perdew}},
  \bibinfo{author}{\bibfnamefont{K.}~\bibnamefont{Burke}}, \bibnamefont{and}
  \bibinfo{author}{\bibfnamefont{M.}~\bibnamefont{Ernzerhof}},
  \bibinfo{journal}{Physical Review Letters} \textbf{\bibinfo{volume}{77}},
  \bibinfo{pages}{3865} (\bibinfo{year}{1996}), ISSN \bibinfo{issn}{0031-9007,
  1079-7114},
  \urlprefix\url{https://link.aps.org/doi/10.1103/PhysRevLett.77.3865}.

\end{thebibliography}

\section*{Acknowledgments}

This work was supported by computational resources provided by the Australian Government through NCI and Pawsey under the National Computational Merit Allocation Scheme. S.A.T and S.R acknowledge computational resources that were received through the Australian Research Council funded Center of Excellence in Exciton Science (CE170100026).


\end{document}